\documentclass[10pt, letterpaper, twocolumn, twoside, conference, final]{IEEEtran}
\usepackage{dsfont}
\usepackage{cite}
\usepackage{graphicx}
\usepackage{amsmath}
\usepackage{amsfonts}
\usepackage{array}
\usepackage{times}
\usepackage{stfloats}
\usepackage{amssymb}
\usepackage{yhmath}
\usepackage{graphics}
\usepackage{textcomp}
\usepackage{amscd}
\usepackage{epsfig}
\usepackage{psfrag}
\usepackage{rotating}
\usepackage{amsmath}
\usepackage{url}
\usepackage{color}
\usepackage{float}
\usepackage{balance}

\begin{document}
\title{Optimal Power Allocation in Block Fading Gaussian Channels with Causal CSI and Secrecy Constraints}
\author{Arsenia Chorti$^{\dag}$,  Katerina Papadaki$^{\ddag}$,  H. Vincent Poor$^*$ \\
\small{$^\dag$School of Computer Science and Electronic Engineering, Wivenhoe Park, Colchester, CO4 3SQ, UK}\\
\small{$^\ddag$Department of Management,
London School of Economics and Political Science,
Houghton Street,
London WC2A 2AE} \\
\small{$^*$Department of Electrical Engineering, EQUAD, 19 Olden  Street, Princeton
University, Princeton, New Jersey 08544, USA}\\
achorti@essex.ac.uk,  k.p.papadaki@lse.ac.uk,  \{achorti, poor\}@princeton.edu}
\maketitle

\begin{abstract}
The optimal power allocation that maximizes the secrecy capacity of block fading Gaussian (BF-Gaussian) networks with causal channel state information (CSI),  $M$-block delay tolerance and a frame based power constraint is examined. In particular, we formulate the secrecy capacity maximization as a dynamic program. We propose suitable linear approximations of the secrecy capacity density in the low SNR, the high SNR and the intermediate SNR regimes, according to the overall available power budget. Our findings indicate that when the available power resources are very low (low SNR case) the optimal strategy is a threshold policy. On the other hand when the available power budget is infinite (high SNR case) a constant power policy maximizes the frame secrecy capacity. Finally, when the power budget is finite (medium SNR case), an approximate tractable power allocation policy is derived.
\end{abstract}


\section{Introduction}

Physical
layer security (PLS) investigates the potential of taking advantage of the impairments in real communication media, such as fading or noise in wireless channels, in order
to achieve confidentiality in data exchange. PLS was pioneered
by Wyner, who introduced the wiretap channel and established the possibility of creating perfectly secure
communication links without relying on private (secret) keys \cite{Wyner75}. Recently, there have been
considerable efforts devoted to generalizing this result to the wireless fading channel and to multi-user
scenarios \cite{Poor08}, \cite{Gopala08},  \cite{Chorti13}.

\par In the present study we investigate optimal power allocation policies in block fading Gaussian (BF-Gaussian) wireless networks with secrecy and delay constrains. In our model, a transmitter wishes to broadcast secret messages to a legitimate user by employing physical layer security approaches, subject to a strict $M$-block delay constraint; accordingly, at the source a stochastic encoder maps the confidential messages to codewords of length $n= M N$ transmitted over $M$ independent blocks, i.e., we assume that an interleaver of at most depth $M$ is employed. We assume that the fading realizations are independent and identically distributed (i.i.d), that they remain constant over each block of $N$ channel uses and that they change independently from one block to the next.

In the investigated setting, in order for random coding arguments to hold it is required that $n \rightarrow  \infty$. For finite $n$, the BF-Gaussian channels are typically not information-stable and the generalized capacity expressions in \cite{Verdu94} need to be employed. In this work, similarly to the work in \cite{Caire99}, we bypass such issues by assuming that $N \rightarrow \infty$. An alternative line of work was suggested in \cite{Gungor11} by jointly employing queues of secret keys allowing for the avoidance of secrecy outage events; however this option is not considered at present. The case of $M\rightarrow \infty$ that corresponds to the ergodic channel has been investigated in \cite{Poor08} and \cite{Gopala08}.

\par The presentation of the present work is organized as follows. First, we restate the secure waterfilling solution to the optimal power allocation optimization problem in $M$-block BF-Gaussian networks with acausal channel state information (CSI). This framework is pertinent to applications with parallel channels (e.g. in the frequency domain) under short-term power constraints (e.g. OFDM networks with frame based power constraints). Assuming that the $M$-block CSI is available at the transmitting and receiving nodes at the beginning of the transmission frame, the secure waterfilling policy that maximizes the network secrecy capacity \cite{ChortiATC13} is discussed.

Next, we investigate BF-Gaussian channels with long term power constraints. We begin with a ``blind scenario'' in which the optimal power allocation is to be decided without any CSI information; the statistics of the channel gains are the only variables in the power allocation decision process. We use this setting to demonstrate that the formulation of the optimal power allocation problem, maximizing the secrecy capacity subject to a delay, as a dynamic program leads naturally to intuitive and analytic solutions. In particular, in absence of any CSI information we show that the optimal policy is to equally distribute the power budget in the $M$ transmission blocks, as long as the expected value of the difference of the channel gains of the legitimate and eavesdropping terminals is positive.

Then, we examine networks with causal access to the legitimate user's and the eavesdropper's CSIs over a horizon of $M$ transmission blocks; the pairs of the legitimate user's and eavesdropper's channel gains are sequentially revealed to the network nodes. We distinguish three subcases accounting for the low, high and intermediate SNR regions. In the low SNR a threshold transmission policy is shown to be approximately optimal, in line with earlier results in networks without secrecy constraints \cite{Cioffi02}. On the contrary, we demonstrate that in the high SNR case, the optimal strategy is to transmit with constant power in those blocks in which a non zero secrecy capacity can be achieved. Finally, for intermediate SNRs we derive a tractable expression for the transmission policy that depends on the gap between the legitimate and eavesdropping receivers' CSIs.

 \section{System Model}
We assume a BF-Gaussian channel with i.i.d. realizations. During the $m$-th transmission block the legitimate user's channel gain is denoted by $\alpha_m$ and the eavesdropper's channel gain by $\beta_m$. We can exploit the fact that BF-Gaussian channels are weakly symmetric \cite{Bloch_Barros11} to simplify the proofs of the coding theorems.

\par \textit{Definition}: The secrecy capacity density during one transmission block of the BF-Gaussian channel for an input power $\gamma$ and channel gains $(\alpha, \beta)$ can be expressed as
\begin{equation}
c_s({\gamma, \alpha, \beta})\doteq {\Bigg[\log\frac{1+\alpha\gamma}{1+\beta\gamma}}\Bigg]^+
\end{equation}
with $[\cdot]^+=\max(\cdot, 0) $.
The secrecy capacity of the $M$-block transmission frame for a vector of input powers $\mathbf{\gamma}=[\gamma_1, \gamma_2, \ldots, \gamma_M]$ and pairs of channel gains $(\mathbf{\alpha, \beta})=[(\alpha_1, \beta_1), (\alpha_2, \beta_2), \dots, (\alpha_M, \beta_M)]$, can be expressed as:
\begin{equation}
C_s\doteq\frac{1}{M}\sum_{m=1}^{M}{c_s(\gamma_m, \alpha_m, \beta_m)}.
\end{equation}
\section{Power Control with Short-term Power Constraint and Full $M$-Block CSI}

The optimal power allocation policy assuming that at the beginning of the transmission frame the CSI of $M$ parallel blocks is revealed to the transmitting and  receiving nodes has been derived in \cite{ChortiATC13} and is repeated below for convenience. This is the baseline secure waterfilling policy and its performance cannot be exceeded in the causal scenario.
 \par  Without loss of generality we assume that the pairs of channel gains $(\alpha_m, \beta_m)$, $m=1, \ldots, M$ are already permuted so that the differences
 \begin{equation}
\delta_m=\alpha_m-\beta_m \label{eq:delta}
\end{equation}
 appear in non-increasing order. The optimal power allocation problem can be stated as:
 \begin{eqnarray}
 &&\max_{\gamma} C_s \\
 &&\text{s.t. }\sum_{m=1}^{M}{\gamma_m}\leq M {P} \text{ and } \gamma_m\geq 0, m=1,\ldots,M.\label{eq:power constraint}
  \end{eqnarray}

 We further define the inverse channel gaps $d_m$ as:
 \begin{equation}
 d_m=\frac{1}{\beta_m}-\frac{1}{\alpha_m}.
 \end{equation}
 The power allocation $\mathbf{\gamma^*}=(\gamma_1^*, \gamma_2^*, \ldots, \gamma^*_M)$ that maximizes the secrecy capacity satisfies the $M$-block power constraint with equality, i.e,
 \begin{equation}
  \sum_{m=1}^{M}{\gamma_m^*}= M {P},
  \end{equation}
  and is given by the secure waterfilling algorithm
 \begin{equation}
 \gamma_m^*\bigg(\frac{1}{\lambda}\bigg)=\left\{\begin{array}{ll}
                                           \frac{1}{2}\Big[ \sqrt{d_m ^2+\frac{4}{\lambda} d_m} -\bigg(\frac{2}{\alpha_m}+d_m\bigg)\Big],& m \in \mathbb{Q} \\
                                           0, & \text{otherwise}
                                         \end{array}\right.
                                          \end{equation}
 where $\mathbb{Q}=\{i: {\lambda}^{-1} \geq {\delta_i}^{-1}\}$.

 The functions $\gamma_m^*({\lambda}^{-1})$ are monotone increasing and continuous in ${\lambda}^{-1}$. As a result, there exists a unique integer 
 $\mu$ in $\{1, \ldots, M \}$ such that $\lambda^{-1}\geq {\delta_m}^{-1}$ for $m\leq \mu$ and $\lambda^{-1} < {\delta_m}^{-1}$ for $m> \mu$. The waterlevel ${\lambda}^{-1}$ can be derived by sequentially pouring water to the functions $\gamma_m^*({\lambda}^{-1})$ until the power constraint is met with equality, i.e., $\sum_{m=1}^{\mu}{\gamma_m^*({\lambda}^{-1})}= M {P}$.

\section{Power Control with Long-term Power Constraint without CSI}
We assume an overall long-term power constraint over $M$ sequential transmission blocks in the form of (\ref{eq:power constraint}).
Accordingly, the channel gains of the legitimate user and the eavesdropper during the $m$-th block are denoted as $\alpha_m$ and $\beta_m$ with known distributions $p_{A}(\alpha)$ and $p_{B}(\beta)$ respectively.
Our objective at block $m$, given that we have remaining power $p_m$, is the identification of the power allocation $\gamma_m^*$ that maximizes the instantaneous secrecy capacity and the secrecy capacity for the future transmission blocks from block $m+1$ to $M$.

\subsection{Blind Scenario}
We first consider the case in which during the $m$-th block we take a decision on the value of $\gamma_m$ without having information on the current channel gains ($\alpha_m,\beta_m)$, except for their distributions and the remaining power $p_m$. In this formulation, our objective is to maximize the \textit{expected}  secrecy capacity over the entire horizon. Let $\gamma = (\gamma_1,\ldots,\gamma_{M})$. The stochastic optimization objective function can be written as follows:
\begin{equation}
\max_{\gamma} \mathbb{E}\Bigg\{ \sum_{m=1}^{M} c_s(\gamma_m, \alpha_m, \beta_m) \Bigg\}=\max_{\gamma} \mathbb{E}\Bigg\{ \sum_{m=1}^{M} c_s(\gamma_m, \alpha, \beta) \Bigg\},
\label{eq:problem_blind_channel}
\end{equation}
where the expectation taken over the random variables $\alpha_m$ and $\beta_m$ is re-written with rapport to the generic random variables $\alpha$ and $\beta$.

The above problem can be written as a stochastic dynamic program as follows: We let $V_m(p_m)$ be the aggregate secrecy capacity density gained from block $m$ to the end of the horizon if the optimal power allocation policy is used. Then the dynamic programming equations can be written as:
\begin{eqnarray}
V_m(p_m) &=& \max_{0 \leq \gamma_m \leq p_m} \mathbb{E}\{c_s(\gamma_m, \alpha, \beta)+V_{m+1}(p_m-\gamma_m)\} \nonumber\\
&&\text{                       } m=1,\ldots,M  \nonumber \\
V_M(p_{M+1}) &=& 0 \text{  (resources exhausted).}
\label{eq:dp_blind_channel}
\end{eqnarray}

We perform backward dynamic programming on the optimality equations (\ref{eq:dp_blind_channel}). We define the function:
\begin{equation}
f(\gamma) \equiv \mathbb{E}\left\{\log\frac{1+\alpha\gamma}{1+\beta\gamma}  \right\}.
\end{equation}
We start the dynamic programming recursion at block $m=M$, where the optimality equations are:
\begin{equation}
V_{M}(p_{M}) = \max_{0 \leq \gamma_{M} \leq p_{M}} f(\gamma_{M})\mathds{1}_{\mathbb{E}\{\alpha-\beta\}>0},
\label{eq:T-1}
\end{equation}
where $\mathds{1}_{\mathbb{F}(\cdot)}$ denotes the indicator function.
Since $f$ is nondecreasing, the maximization in (\ref{eq:T-1}) is achieved at $\gamma_{M}^* = p_{M}$ if $\mathbb{E}\{\alpha-\beta\}>0$ and for any value of the power if $\mathbb{E}\{\alpha-\beta\}<0$. The ambiguity in the latter scenario is resolved by imposing $\gamma_{M}^*=0$ whenever this occurs, i.e.,
\begin{eqnarray}
\gamma_{M}^*=\left \{ \begin{matrix}
p_{M}, & \text{ if } \mathbb{E}\{\alpha-\beta\}>0 \\
0, & \text{otherwise.}
\end{matrix}      \right .
\end{eqnarray}

When the condition $\mathbb{E}\{\alpha-\beta\}>0$ is satisfied, we have $V_{M}(p_{M}) = f(p_{M})$.
In this case, at block $m=M-1$ the optimality equations are:
\begin{equation}
\begin{aligned}
V_{M-1}(p_{M-1})& = \max_{0 \leq \gamma_{M-1} \leq p_{M-1}} f(\gamma_{M-1}) + f(p_{M-1}-\gamma_{M-1}). \label{eq:M-2 blind}
\end{aligned}
\end{equation}
Let $h(\gamma) = f(\gamma) + f(p-\gamma)$. Note that $h'(\gamma) = f'(\gamma) - f'(p-\gamma)$, and since $f'(\gamma)$ is nonincreasing and $f'(p-\gamma)$ is nondecreasing in $\gamma$, we have that $h'$ is nonincreasing. This means that it can have at most one extreme point in the interval $[0,p]$, and the extreme point must be a maximum. At $\gamma = \frac{p}{2}$ we have: $h'\left(\frac{p}{2}\right) = f'(\frac{p}{2}) - f'(\frac{p}{2}) = 0$. Therefore in (\ref{eq:M-2 blind}) the maximum is achieved at $\gamma_{M-1}^* = \frac{p_{M-1}}{2}$ and $V_{M-1}(p_{M-1}) = 2 f(\frac{p_{M-1}}{2})$.

 Continuing the recursion we get
 \begin{equation}
 V_{M-n}(p_{M-n}) = (n+1) f\big(\frac{p_{M-n}}{n+1}\big)
  \end{equation}
  and the optimal decision is $\gamma_{M-n}^* = \frac{p_{M-n}}{n+1}$. This implies that if we have no information about the channel the optimal thing to do is to divide the power into as many equal parts as there are periods remaining, i.e., for $m=1, \ldots,M$ and \textit{any distributions} $p_A(\alpha)$ and $p_B(\beta)$
\begin{eqnarray}
\gamma_m^*=\left\{ \begin{matrix}
\frac{P}{M}, & \text{ if } \mathbb{E}\{\delta\}>0 \\
0, & \text{ otherwise,}
\end{matrix}
\right.
\end{eqnarray}
with $\delta$ defined as the gap of channel gains given in (\ref{eq:delta}).

The above results are intuitive; as expected, the blind maximization of a function of the outcomes of $M$ independent trials can be achieved by equidistribution of the available resources. What is surprising though, is that the results are independent of the statistics of the underlying processes and only require knowledge of the expected value of the gap between the legitimate and eavesdropper's channel gains.

\section{Power Control with Long-term Power Constraint and Causal CSI}
In the current section we investigate the case in which during the $m$-th transmission block we causally obtain information regarding the channel state, i.e., we have access to $(\alpha_m, \beta_m)$. In this setting, during the $m$-th transmission block, we have to solve the optimization problem
\begin{eqnarray}
 V_m(p_m)&=&\max_{0\leq \gamma_m \leq p_m}c_s(\alpha_m, \beta_m, \gamma_m) \nonumber \\
&+& \mathbb{E}\Bigg\{\sum_{n=m+1}^{M}{ c_s(\gamma_n, \alpha, \beta) }\Bigg\}, \\
&\text{s.t. }& \sum_{m=1}^{M}{\gamma_m}\leq M P .
\end{eqnarray}

We distinguish three cases, according to the available power budget $P$; the low SNR, the high SNR and the intermediate SNR cases.

\subsection{Low SNR}
In the low SNR, the available power is assumed small, i.e., $P\ll 1$. As a result a valid linear approximation of the logarithmic function would be $\log(1+z)\simeq z$, leading to an approximate expression for the secrecy capacity density given by:
\begin{eqnarray}
c_s(\gamma, \alpha, \beta)\simeq [\alpha-\beta]^+\gamma=[\delta]^+\gamma,
\end{eqnarray}
with $\delta$ defined in (\ref{eq:delta}).
The function $V_m$ to be optimized at $m=M$ could then be written as
\begin{equation}
V_{M}(p_{M})=\max_{0\leq \gamma_{M} \leq p_{M}} [\delta_M]^+\gamma_{M}.
\end{equation}
The objective is thus approximated as a linear function of the power, so that at $m=M$ the optimal power allocation is straightforwardly given by
\begin{eqnarray}
\gamma_{M}^*=\left \{ \begin{matrix}
p_{M}, & \text{ if  } \delta_{M}>0,\\
0, &\text{ otherwise}
\end{matrix}\right. =[\delta_M]^+p_M.
\end{eqnarray}
At $m=M-1$ the objective function takes the form
\begin{eqnarray}
V_{M-1}(p_{M-1})&=&\max_{0\leq \gamma_{M-1} \leq p_{M-1}}   [\delta_{M-1}]^+\gamma_{M-1} \nonumber\\
&+& \mathbb{E} \{[\delta]^+\}(p_{M-1}-\gamma_{M-1}).
\end{eqnarray}
Thus, at $m=M-1$, the optimal power allocation is given by
\begin{eqnarray}
\gamma_{M-1}^*=\left\{ \begin{matrix}
                    0, & \text{if } [\delta_{M-1}]^+\leq\mathbb{E} \{[\delta]^+\}   \\
                    p_{M-1}, & \text{if } [\delta_{M-1}]^+>\mathbb{E} \{[\delta]^+\}
                  \end{matrix} \right .
\end{eqnarray}

Continuing on the backwards recursion, the optimal power policy during the $m$-th block is derived as:
\begin{eqnarray}
\gamma_{m}^*=\left\{ \begin{matrix}
                    0, & \text{if } [\delta_{m}]^+\leq\mathbb{E} \{[\delta]^+\}   \\
                    p_{m}, & \text{if } [\delta_{m}[^+>\mathbb{E} \{[\delta]^+\}\\
                  \end{matrix} \right . \label{eq: policy low SNR}
\end{eqnarray}
for $m=0,\ldots, M-1$.
In the proposed threshold power policy, whenever a ''good enough" gap in the channel gains $\delta_m$ of the legitimate and the eavesdropping receivers occurs then we transmit at full power.

\begin{figure*}[!b]
\normalsize 
\setcounter{equation}{32}
\hrulefill \vspace*{4pt}
\begin{eqnarray}
V_{M-1}(p_{M-1}) &=& \max_{0\leq \gamma_{M-1}\leq p_{M-1}} \frac{2}{\ln(2)} \frac{[\delta_{M-1}]^+\gamma_{M-1}}{2+(\delta_{M-1}+2\beta_{M-1})\gamma_{M-1}}+ \frac{2}{\ln(2)} \frac{\mathbb{E}\{\delta^2\}(p_{M-1}-\gamma_{M-1})}{2+\mathbb{E}\{(\delta+2\beta)\mathbb{E}\{[\delta]^+\}(p_{M-1}-\gamma_{M-1})} \nonumber\\
\end{eqnarray}
\end{figure*}
\setcounter{equation}{24}

Intuitively, in the low SNR there will not be many opportunities for achieving high values of the secrecy capacity density, so whenever such an opportunity occurs it should be seized in order to maximize the secrecy capacity over the whole horizon.
The threshold is fixed to the expected value of the gap between the channel gains of the legitimate user and the eavesdropper, lower bounded by zero. Even when the legitimate user's channel is on average worse than the eavesdropper's, it is still possible to transmit at some non-zero rate \text{even} in the low SNR, given a long enough horizon, i.e., for $M \gg$.

\subsection{High SNR}
In the high SNR, i.e., for $P \rightarrow \infty$, we can transmit at very high power during any of the transmission blocks.
A good approximation for the secrecy capacity density during the $m$-th block is derived as
\begin{equation}
\lim_{\gamma\rightarrow \infty}c_s(\gamma, \alpha, \beta)= \Bigg[\log\frac{\alpha}{\beta}\Bigg]^+.
\end{equation}
The optimization problem of the secrecy capacity density is as a result independent of the power allocation and any transmission policy could be used. Accounting for other important considerations, e.g. the minimization of the information leakage, it is proposed to only transmit during the blocks that satisfy the condition $\delta_m>0$, i.e.,
\begin{equation}
\gamma_m^*=\left\{  \begin{matrix}
0, & \text{if } \delta_m\leq 0 \\
\frac{p_m}{M-m}, & \text{if } \delta_m>0\\
\end{matrix}
\right . =[\delta_m]^+ \frac{p_m}{M-m}, \label{eq:policy high SNR}
\end{equation}
with $p_1=P$ and $m=1, \ldots, M$.

\subsection{Intermediate SNR}
Aiming at producing tractable expressions for the power allocation, we  propose using the following approximation for the logarithmic function \cite{Abramowitz}:
\begin{eqnarray}
   \ln (z) &=&2\sum_{n=0}^{\infty}{\frac{1}{2n+1}\Big(\frac{z-1}{z+1}\Big)^{2n+1}} \Rightarrow \nonumber\\
   \log(z)&\simeq& \frac{2}{\ln(2)}\frac{z-1}{z+1}. \label{eq:log approx}
\end{eqnarray}
In addition, we will use the following first order approximation:
for the correlated random variables $X$ and $Y$ and linear functions $F(\cdot)$ and $G(\cdot)$,
\begin{equation}
\mathbb{E}\Bigg\{ \frac{F(X)}{G(Y)}  \Bigg\} \simeq \frac{\mathbb{E}\{F(X)\}}{\mathbb{E}\{G(Y)\}}. \label{eq:ratio approx}
\end{equation}

Using (\ref{eq:log approx}), the secrecy capacity density can be expressed as:
\begin{equation}
c_s(\gamma, \alpha, \beta)\simeq \frac{2}{\ln(2)}\frac{[\delta]^+\gamma}{2+(\delta+2\beta)\gamma} ,
\end{equation}
while employing (\ref{eq:ratio approx})
\begin{equation}
\mathbb{E}\{c_s(\gamma, \alpha, \beta)\}\simeq \frac{2}{\ln(2)}\frac{\mathbb{E}\{[\delta]^+\}\gamma}{2+\mathbb{E}\{(\delta+2\beta)\}\gamma} .
\end{equation}
Notably, the quantity $\delta$ defined in (\ref{eq:delta}) of the acausal CSI secure waterfilling case reappears in the optimization problem.

Performing backward dynamic programming on the optimality equations we get:
\par $\rhd$ At block $m=M$,
\begin{eqnarray}
V_{{M}}(p_{M})&=&\max_{0\leq \gamma_{M}\leq p_{M}} \frac{2}{\ln(2)} \frac{[\delta_M]^+\gamma_M}{2+(\delta_M+2\beta_M)\gamma_M} \Rightarrow \nonumber\\
\gamma_M^*&=&[\delta_M]^+p_M, \\
\text{while for } p_M&=&\gamma_M^*,  \text{ we get} \nonumber\\
V_M(p_M)&=&\frac{2}{\ln(2)} \frac{\delta_M^2p_M}{2+(\delta_M+2\beta_M)[\delta_M]^+p_M}.
\end{eqnarray}

$\rhd$ For the block $m=M-1$, the objective equation is given at the bottom of the page. We note that the objective function is concave in the interval $[0, p_{M-1}]$.
\setcounter{equation}{33}
The optimal policy is as a result derived as
\begin{eqnarray}
\gamma_{M-1}^*&=&
\rho_{M-1}
\end{eqnarray}
with $\rho_{M-1}$ being the positive root of the quadratic equation $\frac{\partial V_{M-1}}{\partial\gamma_{M-1}}=0$, that can be evaluated in analytic form.

Although the analytic expression for the quantity $V_{M-1}(\gamma_{M-1}^*)$ is overly complicated to be included at present, we note that its expectation is easier to evaluate.
%

$\rhd$ Continuing backwards we get that at $m=M-2$ the objective function is also concave in the interval $[0, p_{M-2}]$. 
The optimal policy is derived as
\begin{eqnarray}
\gamma_{M-2}^*&=&
\rho_{M-2},
\end{eqnarray}
with $\rho_{M-2}$ being the positive root of the equation $\frac{\partial V_{M-2}}{\partial\gamma_{M-2}}=0$.
Generalizing we find that the optimal policy is expressed as:
\begin{eqnarray}
\gamma_{m}^*&=&
\rho_{m},  \label{eq:policy medium SNR}
\end{eqnarray}
where $\rho_m$ is the positive root of the equation $\frac{\partial V_{m}}{\partial\gamma_{m}}=0$.

\subsection{Semi-blind Scenario}
The derived results apply also in the semi-blind scenario in which only the legitimate user CSI is causally made available to the transmitting and receiving nodes by substituting $\delta$ by $\alpha$ in the equations (\ref{eq: policy low SNR}), (\ref{eq:policy high SNR}) and (\ref{eq:policy medium SNR}). However, in this case the minimization of the probability of secrecy outage should in principle be investigated instead of the maximization of the secrecy capacity density. At present this is left as future work, along with the numerical evaluation of the performance of the outlined policies.

\section{Conclusions}
In the present work we investigate the optimal power allocation in delay constrained $M$-block BF-Gaussian networks. By studying the blind case with no CSI availability during the decision process we conclude that the optimal policy consists in equally distributing the power along the transmission blocks as long as a positive gap between the channel gains of the legitimate user and the eavesdropper is anticipated. Furthermore, the study of networks with causal access to the CSI was performed accounting for three distinct cases; the low, high and medium SNR. In the low SNR we derived a near optimal threshold policy whereas in the high SNR a constant transmission policy is shown to be optimal. Finally, for intermediate values of the SNR we derive an approximate, analytically tractable dynamic program of reduced computational complexity.

\balance
\bibliographystyle{IEEEtran}
\bibliography{isit14_bib}

\end{document}